\documentclass[a4paper]{iopart}

\newcommand{\solm}{M_{\odot}}

\usepackage{harvard}
\usepackage{graphicx}
\usepackage{subfig}

\begin{document}
\bibliographystyle{jphysicsB}

\title{Central Stars of Planetary Nebulae in SDSS and IPHAS}

\author{S. Weston$^1$, R. Napiwotzki$^1$, S. Sale$^2$}

\address{$^1$Centre for Astrophysics Research, %
University of Hertfordshire, College Lane, Hatfield, Herts, AL10 9AB, UK}
\address{$^2$Imperial College London, Blackett Laboratory, %
Exhibition Road, London, SW7 2AZ, UK }
\ead{s.1.weston@herts.ac.uk}

\begin{abstract}
Space densities and birthrates of Planetary Nebulae (PNe) are highly
uncertain. A large range of formation rates has been derived by
different studies, which has led to contradicting ideas for the final
evolutionary phases of low and intermediate mass stars. We started a
project to deduce a birthrate using a sample of PNe within 2kpc. The
central stars will be identified in the PNe fields by their
photometric colours and then used to establish improved distance
estimates. To facilitate this we have created grids of photometric
colours which are used to constrain stellar parameters. Our study has
concentrated on PNe in SDSS and the INT Photometric H$\alpha$ Survey
(IPHAS) so far. IPHAS is a nearly complete northern galactic plane
survey in H$\alpha$, r' and i' bands. Many previously unknown PNe have
been discovered with IPHAS. We investigate implications of a more
complete local sample on PN birthrate estimates.

\end{abstract}

\section{Introduction}

Planetary Nebulae (PNe) are thought to be the final evolutionary phase
of most low/intermediate mass ($0.1\solm <M<8\solm$) stars as they
leave the asymptotic giant branch (AGB) and evolve onto the white
dwarf (WD) cooling sequence.  PNe are an important tool for an
understanding of the final phases of stellar evolution.\\\\ A hot
topic of discussion is whether PNe are formed mainly (or solely) by
single or binary stellar systems. In the standard single star scenario
the PN is formed when the star leaves the AGB after heavy mass
loss. The ejected envelope lights up as the central star contracts and
increases in temperature. Alternatively, \citeasnoun{mar05} and
\citeasnoun{moe06} suggest that PN are largely created by binary
stars. They argue that if the single star scenario is dominant then
too many PNe will be formed in our galaxy compared to the number
observed, even considering corrections for incompleteness. In the
binary scenario, closely orbiting binaries will come into contact and
orbital energy lost to the common envelope. The energy is used to
eject the surrounding gas and drive the expansion of the
nebula. Further support for a dominant binary channel comes from the
non-spherical morphology of most PNe \cite{nor07}. \\\\ A complete
population census of PNe can be used to compare observations with
theoretical predictions resulting from evolutionary time scales and
birthrates. Current investigations predict a range of formation rates
$(0.2~-~8.0)~\times~10^{-12}~\rm{pc}^{-3}~\rm{yr}^{-1}$
\cite{sok06,ish87}. As each PN forms a WD, the WD birthrate should be
equal or greater than the PNe formation rate. The WD birthrate of
$1.0\pm0.25 \times 10^{-12} \rm{pc}^{-3} \rm{yr}^{-1}$ \cite{lie05}
poses a real problem for many PNe estimates which exceed
this. \citeasnoun{nap01} states that one reason for very high
estimates of PNe space densities and birthrates are underestimates of
PNe distances. \\\\ We intend to improve the local density estimate by
collating known and newly discovered PNe within 2kpc. An important
task is to ensure we only include PNe within our volume limited sample
which requires a distance estimate. We will obtain distances by using
the central star of the PN (CSPN). Therefore, we must identify the
CSPN within the PN field as not all are located in the centre. We
discuss locating the central star in
Section~\ref{sec:locate_cspn}. Two methods for determining distance to
CSPNe are explained in Section~\ref{sec:distance}. Finally, we give
details of model grids which will improve the locating the central
star method previously mentioned as well as give stellar parameters
using photometry in Section~\ref{sec:model_grids}.

\section{Locating the Central Star} \label{sec:locate_cspn}
Central stars have unique photometric colours which distinguish them
from most other stellar objects, so a semi-emprical selection region
was defined for the whole range of possible CSPNe parameters. The
region was constructed using synthetic photometry of OB stars
\cite{fit05} and DA WDs \cite{hol06} extended to hotter temperatures
and adjusted to find known CSPNe in SDSS. \citeasnoun{fit05} produce
Johnson photometry so we convert it to SDSS. The result positively
showed that most CSPN fall within the defined region (with the
remaining either saturated or their colours distorted by a companion)
and no contaminate objects are detected as possible central
stars. This gives us confidence, at least with no or little reddening,
that CSPN are well defined with four colours and can be identified
within a PN field.

\section{Distance Estimates} \label{sec:distance}

Using the central stars we apply two ways to estimate
distances. Candidate local PNe are rejected or confirmed from our
sample based on the results, or inspected further if close to our distance
limit.

\subsection{Distance Estimates from Evolutionary Tracks}

\begin{figure}
    \subfloat{
      \addtolength{\hoffset}{-1.5cm}
    \begin{minipage}[b]{0.45\linewidth}
       \includegraphics[width=2.6in]{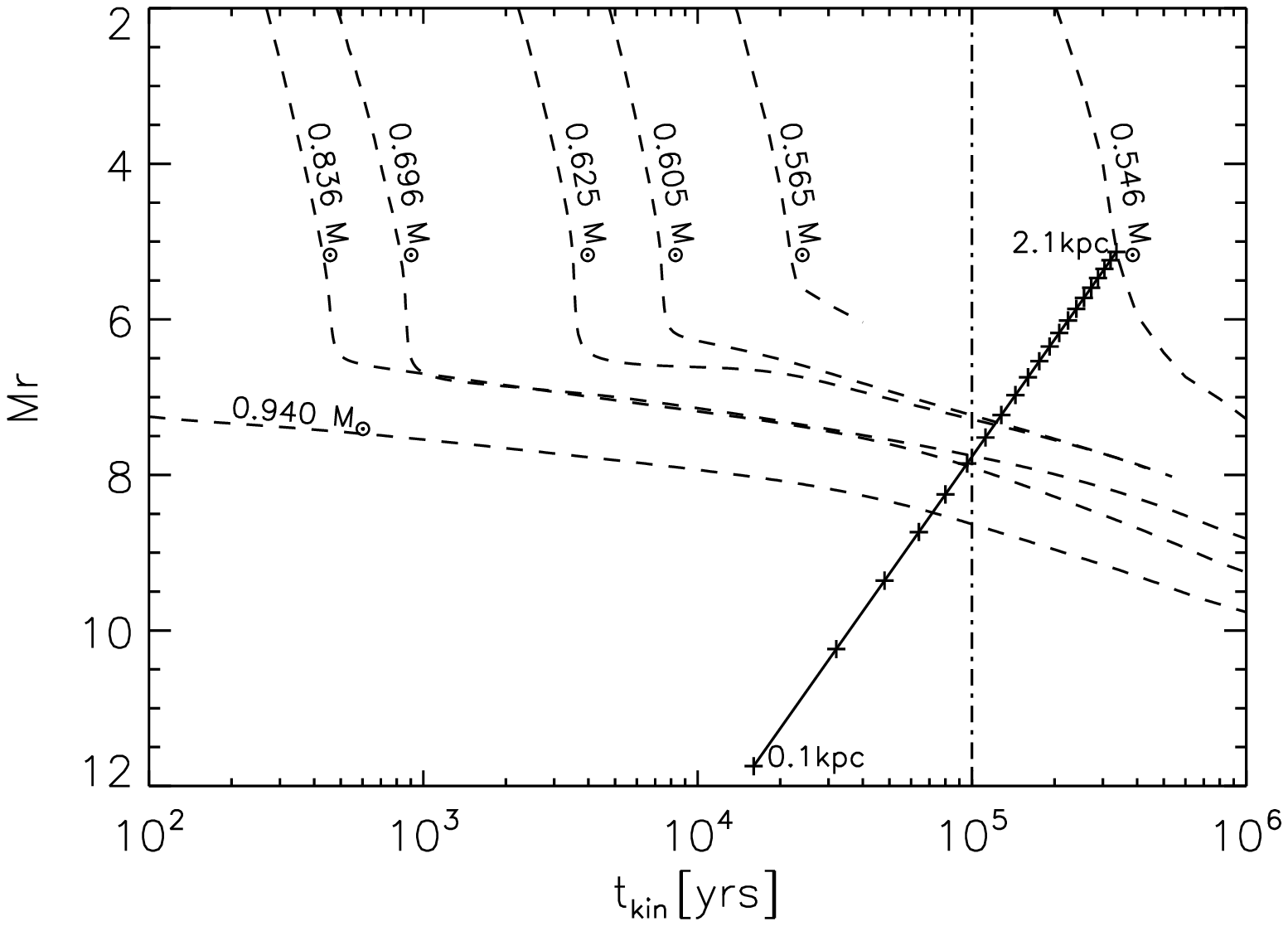}
    \end{minipage}}%
    \subfloat{
    \begin{minipage}[b]{0.45\linewidth}
       \includegraphics[width=2.6in]{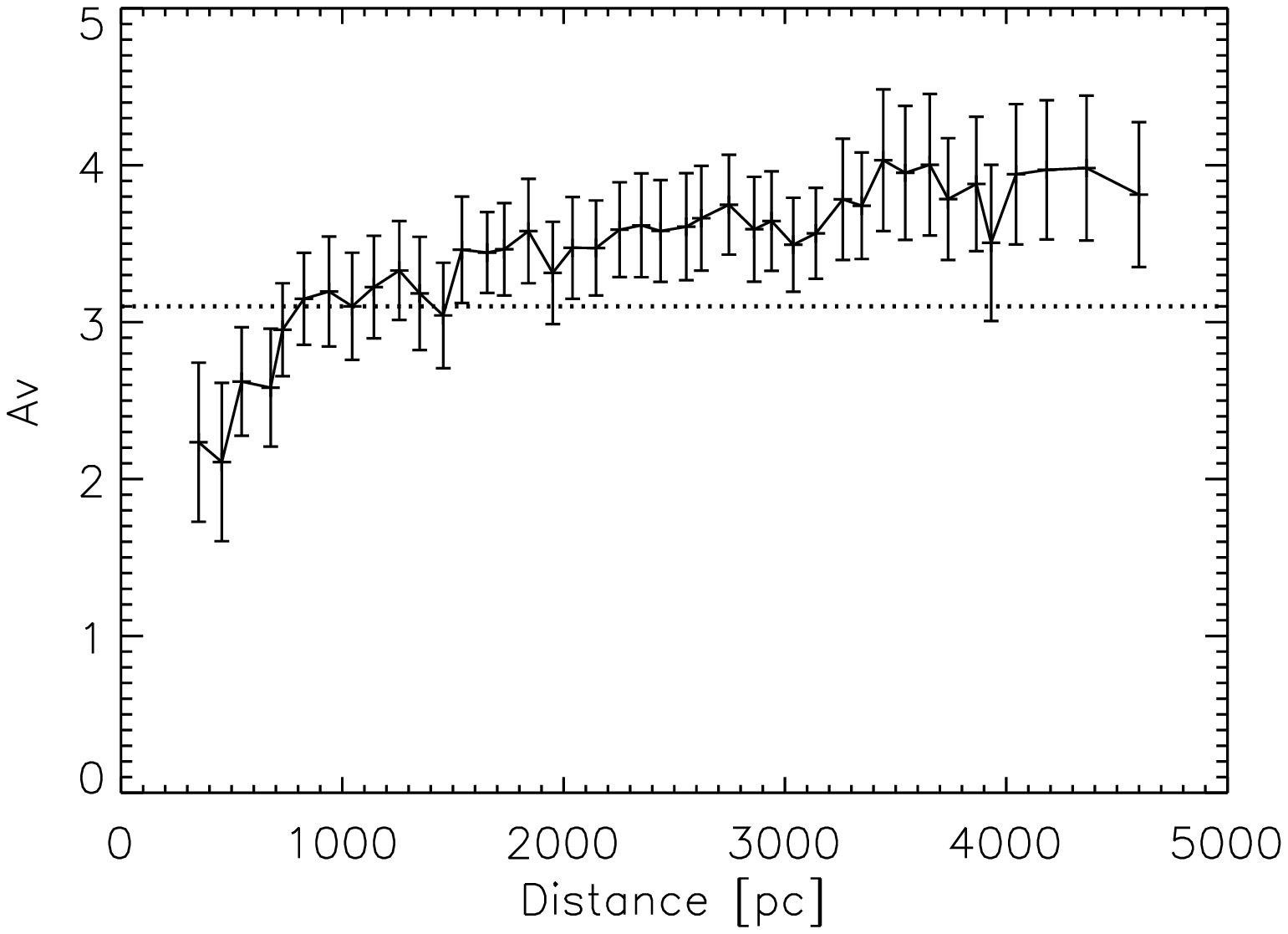}
    \end{minipage}}
    \caption{\small \textit{Left:} PN G158.8+37.1 as an example
      $M_{r'}-t_{kin}$ plot with the evolutionary tracks (dashed
      lines) of Sch\"onberner (1983) and Bl\"ocker (1995). The
      dash-dot line indicates the time scale limit of 100,000 years. The
      resulting upper limit of the PN distance is
      0.6kpc. \textit{Right:} Example of a PN (PN G126.6+01.3) in
      IPHAS and the projected line of sight extinction. This PN lies
      between 0.7 and 4.0 kpc away.}
    \label{fig:distance}     
\end{figure}

\noindent For a PNe with the distance known, an absolute magnitide of
the CSPN and a kinematic age, $t_{\rm{kin}}$, can be computed from the
nebula radius and expansion velocity. Fig. \ref{fig:distance} (left)
shows the post-AGB evolutionary tracks of \citeasnoun{sch83} and
\citeasnoun{blo95} converted to the $M_{r'}-t_{\rm{kin}}$ plane. This
plot can be used to constrain the distance of PNe. For any
hypothetical distance $M_{r'}$ and $t_{\rm{kin}}$ lie on a line, each
point corresponding to a CSPN mass.  We assume 0.6$\solm$\ as the
standard value as 80\% of all white dwarfs have $M=0.6\pm0.1$$\solm$
\cite{lie05}. The angular diameters and expansion velocities are taken
from \citeasnoun{ack92}. Where the expansion velocity is unknown a
value of 20 kms$^{-1}$ is assumed. Extinction is calculated using
$E(r'-i')$ assuming an intrinsic $r'-i'$ value of $-$0.35. \\\\ We
compared our results to distances from \citeasnoun{cah92} and
\citeasnoun{ack92}. The results are mixed with some distances in
agreement, however, most have large discrepancies. Although some
uncertainty results from the the unknown CSPN mass, the statistical
methods used by \citeasnoun{cah92} and \citeasnoun{ack92} have large
systematic errors \cite{nap01}.

\subsection{Distance Estimates from Extinction}
Distances can be estimated using relations between interstellar
absorption and distance. A good distance estimate can only be obtained
with a detailed 3D dust map. The maps of \citeasnoun{sch98} give only
the integrated extinction and thus can't be used for this purpose. Our
future sample will include observations from the INT Photometric
H$\alpha$ survey (IPHAS, Drew et al., 2005\nocite{dre05}). We will
exploit the work of \citeasnoun{sal08} who present an algorithm which will
produce a 3D extinction map across the entire Northern Galactic
plane (Sale et al., in prep.). All CSPNe have a narrow range of
intrinsic $r'-i'$ colours ($r'-i'=-0.35\pm0.04$) and so we can
estimate an accurate $E(r'-i')$. Once converted to an $r'$ band
extinction, $A_{r'}$, the distance can be extracted from Sale et al.'s
projection of the galactic dust. An example can be seen in Fig.
\ref{fig:distance} (right). Combining both methods, we will be able to
compile a sample of good candidates for the local PN population.

\section{Grids of Synthetic Photometry for CSPNe} \label{sec:model_grids}

\begin{figure}
   \subfloat{
    \begin{minipage}[b]{0.45\linewidth}
       \centering \includegraphics[width=2.3in]{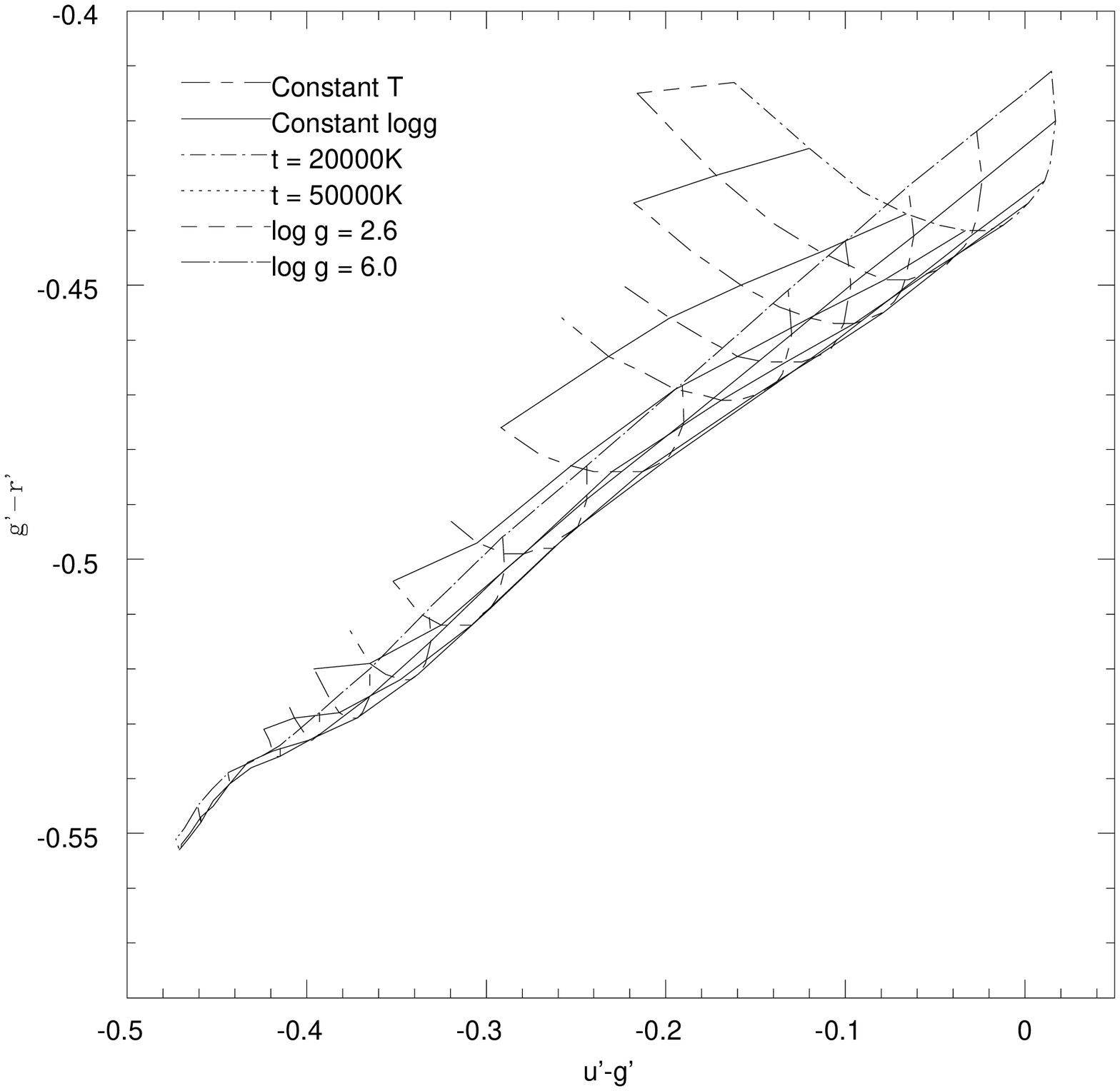}
    \end{minipage}}%
    \hfill
    \subfloat{
    \begin{minipage}[b]{0.45\linewidth}
       \centering \includegraphics[width=2.3in]{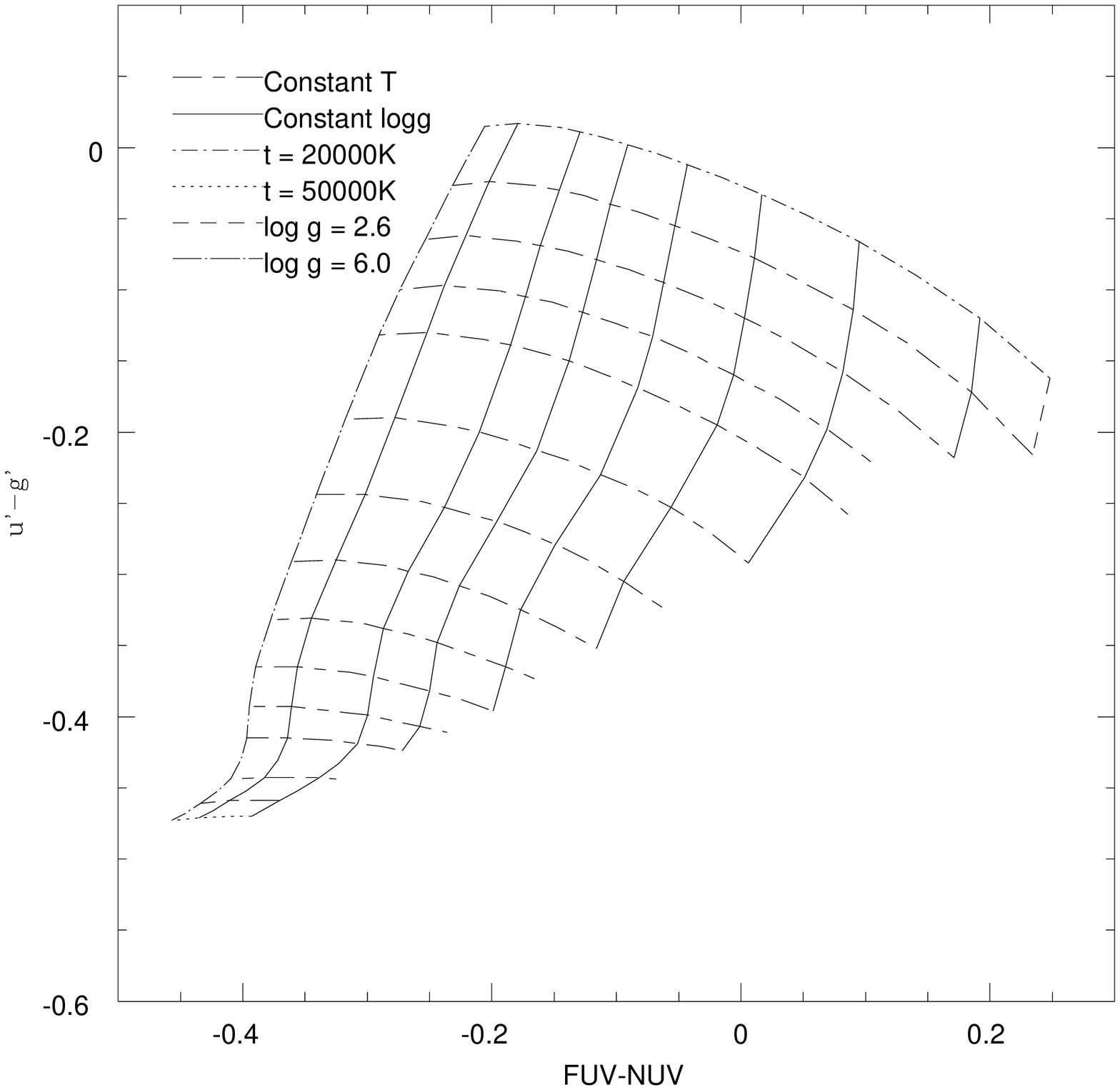}
    \end{minipage}}
    \caption{\small \textit{Left:} CSPN model grid in SDSS
      colour-colour space. These and other optical colour-colour
      diagrams have little sensitivity to log$g$. \textit{Right:} CSPN
      model grid in GALEX--SDSS colour-colour space. GALEX UV bands
      and $u'-g'$ combined are sensitive to temperature and log$g$.}
    \label{fig:modelgrid}     
\end{figure}

We are in the process of refining the photometric selection and
analysis by computing synthetic photometry in all filters required
from model Spectral Energy Distributions (SEDs). This will be extended
to include the whole parameter range for CSPNe and filters covering the
UV to near-IR range. So far we have created photometric grids for OB
type CSPNe from ATLAS9 models \cite{kur91} and hot DA WDs from
\citeasnoun{fin97}. If the star was observed with the UV satellite
GALEX as well, then we can easily determine stellar parameters from
photometry (Fig. \ref{fig:modelgrid}). With the current model
grids for SDSS, GALEX and IPHAS we will be able to locate
the central star, an approximate distance (or range) and stellar
parameters. Furthermore, with the SEDs many bands and systems
(therefore surveys) can be added and the project begin to compile an
accurate and more complete local sample.

\section*{References}
\bibliography{cspn_barca_poster}

\end{document}